\documentclass[showpacs,preprintnumbers,10pt]{revtex4}
\usepackage{amsmath}
\usepackage{graphicx}
\usepackage{amssymb}
\usepackage{floatflt}
\usepackage{epsfig}
\usepackage{epsf}
\usepackage{color}

\newcommand\be {\begin{equation}}
\newcommand\ee {\end{equation}}
\newcommand\bea {\begin{eqnarray}}
\newcommand\eea {\end{eqnarray}}
\newcommand\n {\nonumber}
\newcommand\bc {\begin{center}}
\newcommand\ec {\end{center}}
\newcommand\bfl{\begin{flushleft}}
\newcommand\efl{\end{flushleft}}
\newcommand\bfr{\begin{flushright}}
\newcommand\efr{\end{flushright}}

\begin{document}

\setlength\hoffset{-0.7 cm}
\setlength\voffset{-0.7 cm}
\setlength\evensidemargin{0 cm}
\setlength\oddsidemargin{-0.2 cm}
\setlength\textheight{25cm}

\title { Non-Universal Finite Size Scaling of Rough Surfaces}
\author{Pradipta Kumar Mandal}
\author{Debnarayan Jana}

\email{djphy@caluniv.ac.in}

\affiliation{Department of Physics, University of Calcutta, 92
  A.P.C. Road, Kolkata- 700 009, India}

\date{\today}

\begin{abstract}
We demonstrate the non-universal behavior of finite size scaling in
(1+1) dimension of a nonlinear discrete growth model involving
extended particles in generalized point of view. In particular, we
show the violation of the universal nature of the scaling function
corresponding to the height fluctuation in (1+1) dimension. The 2nd
order moment of the height fluctuation shows three distinct crossover
regions separated by two crossover time scales namely, $t_{\times1}$
and $t_{\times2}$. Each regime has different scaling property. The
overall scaling behavior is postulated with a new scaling relation
represented as the linear sum of two scaling functions valid for each
scaling regime. Besides, we notice the dependence of the roughness
exponents on the finite size of the system. The roughness exponents
corresponding to the rough surface is compared with the growth rate or
the velocity of the surface.

\end{abstract}

\pacs {68.35.Ct, 64.60.al, 61.43.Hv, 05.40.-a}

\maketitle

\section{Introduction}

Theoretical and experimental study of morphology of the growing
surfaces and interfaces, which are generated from various different
growth processes, is a major challenge in recent years
\cite{ba,m,hz}. Several discrete models have been proposed to study
the morphology of the rough surfaces formed in different growth
processes such as imbibition in porous media \cite{adr,abh,kj}, thin
film growth in molecular beam epitaxy (MBE)\cite{asv,k} and growth of
bacterial colonies\cite{bst}. These discrete models are based on
simple stochastic growth rules, such as aggregation and diffusion.

The rough surface evolved from a non-equilibrium process can be
characterized from the study of the moments of the average of height
fluctuations, called surface width, defined as
\be
\label{eq:no1}
W_\gamma(L,t) \,=\, \left[\frac{1}{L}\sum_{i=1}^L  [h(i,t)-\langle h(t)\rangle]^\gamma \right ]^{1/\gamma}
\ee

Where $L$ is the system size, $\langle h(t) \rangle = \frac{1}{L}
\sum_{i=1}^Lh(i,t) $ is the average height over different sites at
time $t$ and $\gamma$ is the order of the moment of the height
fluctuation.

 The dynamic formation of the rough surfaces can be realized with the
 numerical simulation of discrete models. The dynamic and saturated
 behaviors of the rough surface are characterized with the power law
 nature of the moments of the height fluctuation. The short time
 height fluctuation is characterized by the exponent $\beta$ known as
 dynamic one. The saturated behavior of the surface width determines
 the fractal dimension (represented in terms of roughness exponent
 $\alpha$) of the rough surface. The overall scaling behavior of the
 rough surface is studied with the Family-Vicsek scaling ansatz
 \cite{fv}. $W_2(L,t)$ being the 2nd order moment of the height
 fluctuation, one can write the Family-Vicsek phenomenological scaling
 ansatz as

\be
\label{eq:no2}
W_2(L,t)=L^\alpha f \left (\frac{t}{L^z}\right )
\ee

Where  $z = \alpha/\beta$ and $f(u) \sim u^\beta \,\,\,[u \ll 1]$,
$f(u) \sim $ constant \,\, $[u \gg 1]$. This scaling relation is the
central quantity of interest to study the morphology of the surface
for any discrete growth process. The values of the exponents $\alpha$
and $z$ uniquely determine the universality classes of the kinetic
roughening process. Linear discrete model such as random deposition
with surface relaxation (RDSR) \cite{f} and nonlinear discrete models
like  ballistic deposition (BD) model \cite{v}, Eden growth (ED) model
\cite{ed}, restricted solid-on-solid (RSOS) model \cite {kk} and
body-centered solid-on-solid (BCSOS) model \cite{cn} have been
proposed to study the kinetic roughening of the growth processes which
belong to different universality classes. A competitive growth model
consisting of two kinds of particles viz. BD (with probability $1 -p$)
and RDSR (with probability $p$) was reported \cite{cr}. Two system
dependent crossover time scales $t_c$ and $\tau$ were found to
characterize the rough surface. In this model, scaling properties
follow RDSR model for $t \ll t_c$ while for $t \gg t_c$ BD scaling
properties are dominant over the concerned regime. The nonlinear
coupling ($\lambda$) is scaled with the abundance of the particles as
$\lambda \sim p^\gamma$. The overall scaling was found to be

\bea
\label{eq:no3}
W_2(L,t) \sim t^{\beta_{RDSR}}  \,\,\,\,\,\,\,\,\,\,\,\,\,\,\,\,\,\,\,\,\,\,\,\,\,\,\,\,\,\,\,\,\,\,\,\,\,\,\,\,\,\,\,\, t \ll t_c \n \\
W_2(L,t) \sim \lambda^{\beta_{BD}} t^{\beta_{BD}} \,\,\,\,\,\,\,\,\,\,\,\,\,\,\,\,\,\,\,\,\,\,\,\,t_c \ll t \ll \tau \n \\
W_2(L,t) \sim (C_1 + C_2\,p^{3/2})L^\alpha \,\,\,\,\,\,\,\,\,\,\,\,\,\,\,\,\,\,\,\, t \gg \tau
\eea
The suffixes correspond the respective models. $C_1$ and $C_2$ are
constants. $\alpha$ is same for both the models.

Several continuum models have also been prescribed to characterize the
rough surfaces formed due to many natural growth processes
\cite{mmtb}. Continuum linear growth model described by
Edward-Wilkinson (EW) equation \cite {ew} belongs to the same
universality class that of the discrete RDSR model. The values of
$\alpha$ and $\beta$ in (1+1) dimension for this universality class
are given as $\alpha = 1/2$ and $\beta = 1/4$. The nonlinear
Kardar-Parisi-Zhang (KPZ) equation \cite{kpz} defines the universality
class which includes the discrete growth models, such as BD model, ED
model, RSOS model and BCSOS model. In $(1+1)$ dimension the scaling
exponents for this class are $\alpha = 1/2$ and $\beta=1/3$ with the
scaling identity

\be
\label{eq:noref}
\alpha+\alpha/\beta=2
\ee

 This scaling identity follows due to Galilean invariance of the
 interface \cite{mhk}. However, nonlocality in the form of long range
 interaction in KPZ equation can modify the above identity
 \cite{mb,jkk,c}.

Instead of strong agreement between the theoretical and numerical
predictions of $\alpha$ and $\beta$ in the same universality class,
the results do not match with several experimental findings. In (1+1)
dimension, experiments on immiscible fluid displacement show the
roughness exponent $\alpha$ lying between $0.73$ and $ 0.89$
\cite{rdg,redg,hfv1,hfv2}. Besides, experiment on growth of bacterial
colonies \cite{vch} yields $\alpha=0.81$. A possible explanation of
the above mentioned experimental results was proposed by Zhang
\cite{z}. He suggested a model with the noise amplitude having power
law distribution as $P(\eta(\vec{r},t)) \sim \eta^{-(1+\mu)}$, where
$\eta(\vec{r},t)$ is delta-correlated noise.

During recent decades, the growth of organic crystals on inorganic
substrates draws great attention towards the application in electronic
devices \cite{sch}. Thin film growth of several organic crystals on
different substrates \cite{kswm, fwsm, ksd} have been studied
experimentally. One diffusion-limited-aggregation (DLA) type of
discrete growth model \cite{zsg} has been proposed to study the
kinetic roughening of the growth of pentacene film on $SiO_2$
substrate. The growth of organic crystals involves particles of large
sizes, viz, the molecular formula of pentacene, copper phthalocyanin
and 3,4,9,10-perylenetetracarboxylic dianhydride are $C_{22}H_{14}$,
$C_{32}H_{16}N_8Cu$ and $C_{24}H_8O_6$ respectively.

 Motivated by the above various growth mechanisms which involve small
 and extended particles with different types of relaxation rules, we
 have proposed a discrete growth model in (1+1) dimension. This model
 may involve particles of different sizes in a single growth
 process. In this model, the particle size plays a major role in the
 kinetic roughening of the surface. This model shows a morphological
 transition from multifractal to unifractal regime beyond a system
 dependent characteristic length scale \cite{mj}. The morphology of
 the surface depends on the dominating particles in the growth
 process. The finite size scaling relation is not of the universal
 type defined in equation (\ref{eq:no2}),  not even satisfying the
 equation (\ref{eq:no3}). A new scaling relation is proposed to
 characterize the rough surface. The values of the scaling exponents
 $\alpha$ and $\beta$ are not universal for this model. It has a
 finite size dependence with a particular scaling form.

The paper is organized as follows. In Sec II we describe the model
with the simulation parameters. In Sec III, we propose the scaling law
characterizing the rough surface. The effect of finite size on the
scaling exponents is discussed in section IV. Porosity, a parameter
that represents the bulk nature of the system, is introduced in a new
scaling form in section V. Finally, in Sec VI the conclusions are
drawn from the numerical results described in previous sections.

\section{The model and its simulation}

The model introduced here contains essentially the aggregation and
diffusion mechanisms with a new idea of extended particles from a
generalized point of view. The concept of these participating
particles is introduced in such a way so that the particles can be of
various sizes for a single growth process. The diffusion mechanism
considered here is same for all types of particles. In spite of the
same diffusion mechanism  for all particles, however the final
morphology of the surface is not the same when particles of different
sizes are taking part in the growth process separately. A detailed
description of the model is given below.

\begin{figure}[ht]
\begin{center}
{\epsfxsize=8.5cm \epsfysize=5cm \epsfbox{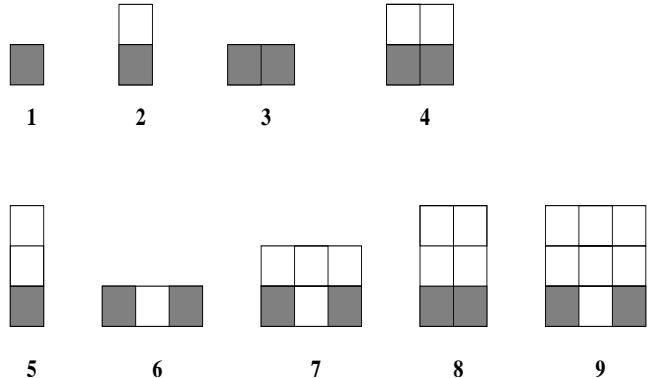}}
\end{center}
\caption{ Different types of particles participating the growth
  process. (1) is the $1 \times 1$ particle, (2) is the $2 \times 1$
  particle, (3) is the $1 \times 2$ particle and so on. The shaded
  boxes of each particle are the extreme cells as defined in the
  text. }
\label{fig:no1}
\end{figure}

The substrate lattice is taken of size $L$ with periodic boundary
condition. The participating particles in this model are considered as
different multiple matrix sequence in terms of the smallest unit of
the substrate lattice. In this way, if the substrate lattice is
considered as a $1 \times L$ matrix such that the smallest unit of
this lattice is a $1 \times 1$ matrix , then the involving particles
may be of sizes as $1 \times 1$ (\raisebox{1.2
  ex}{\setlength\fboxrule{0.8pt}\setlength\fboxsep{1.5mm}\fbox{}}), $1
\times 2$ (\raisebox{1
  ex}{\setlength\fboxrule{0.8pt}\setlength\fboxsep{1.5mm}\fbox{\hspace{0.3
      cm} }} $\equiv$ \raisebox{1.2
  ex}{\setlength\fboxrule{0.8pt}\setlength\fboxsep{1.5mm}\fbox{}\setlength\fboxrule{0.8pt}\setlength\fboxsep{1.5mm}\fbox{}}), $2 \times 1$
(\rotatebox{90}{\setlength\fboxrule{0.8pt}\setlength\fboxsep{1.5mm}\fbox{\hspace{0.22 cm} }} \raisebox{1.2 ex} { $\equiv$ } 
\rotatebox{90}{\setlength\fboxrule{0.8pt}\setlength\fboxsep{1.5mm}\fbox{}\setlength\fboxrule{0.8pt}\setlength\fboxsep{1.5mm}\fbox{}}\,)and
so on. From this generalized point of view, we call the particles as
`extended particle'. The units of $1 \times 1$ particles at the ends
of each extended particles which are facing towards the substrate
lattice will be called `extreme cells'. In this sense, the particles
$1 \times 1$, $2 \times 1$, $3 \times 1$, $.....$ have only one
extreme cell while the particles $1 \times 2$, $2 \times 2$, $1 \times
3$, $......$ have two extreme cells (see Fig.\ref{fig:no1}, extreme
cells are shown as shaded boxes).

When these above mentioned particles form the rough surface, a `stable
position' should be maintained throughout the surface for each of the
involving particles. The stable position refers to a condition when at
least one point from each of the extreme cells for each extended
particles get contact with the columns from the substrate lattice
separately. In other words, the extended particles will attach with
the substrate lattice if at least one corner from each extreme cell is
shared with the corners of the columns from the substrate lattice. In
this circumstance, the selection of the cell from the two extreme
cells by the substrate lattice site is random. The overall aggregation
and diffusion mechanism of the different types of particles is
described as follows: one of the extreme cells of each of the extended
particles is chosen randomly by a substrate site, then it (the
extended particle) slide according to the lower height profile of the
nearest-neighbor columns with a stable position. The process continues
till the stable position is reached. Naturally, when the the number of
extreme cells is one, then the selection of this extreme cell by the
substrate lattice site is completely deterministic.

\begin{figure}[ht]
\begin{center}
{\epsfxsize=7.5 cm \epsfysize=7.5cm \epsfbox{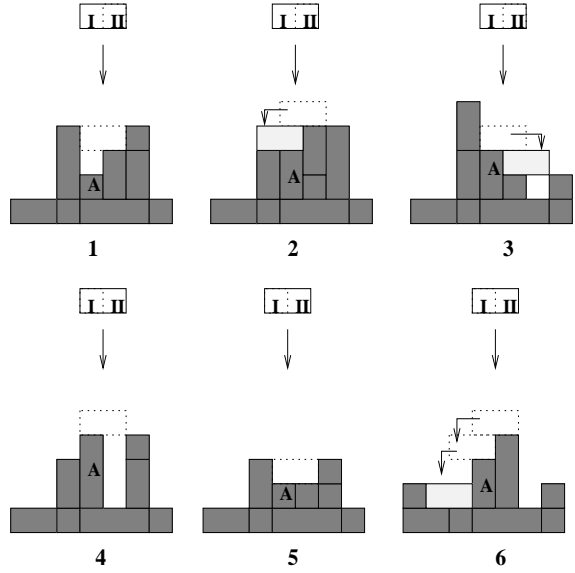}}
\end{center}
\caption{ The aggregation and diffusion mechanism for $1 \times 2$
  particles. Different possible cases are shown. Here one of the
  extreme cell (I) is chosen by a site (A) of the substrate lattice
  randomly. The case will be equivalent when the other extreme cell
  (II) will be chosen by the site (A) of the substrate lattice.}
\label{fig:no2}
\end{figure}

\begin{figure}[ht]
\begin{center}
{\epsfxsize=7 cm \epsfysize= 9.5cm \epsfbox{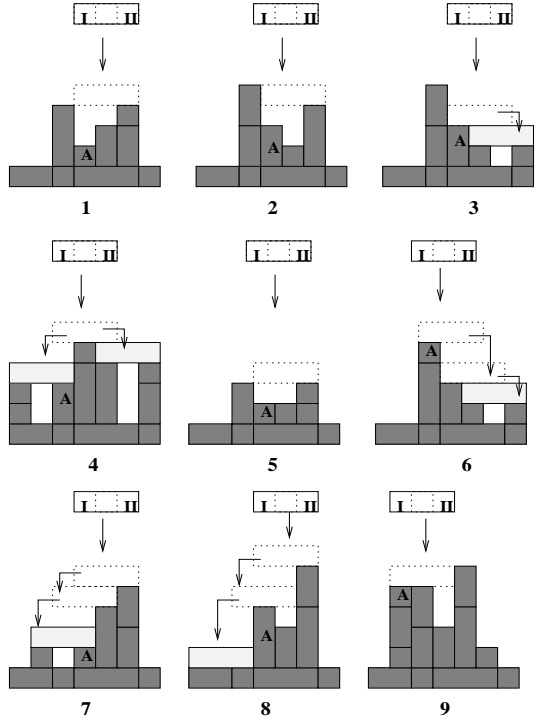}}
\end{center}
\caption{ The deposition and relaxation mechanism for the $1 \times 3$
  particles. Different possible cases are shown. A random lattice site
  (A) chooses one of the extreme cells (I) of the extended $1\times 3$
  particle randomly. The choice of another extreme cell (II) give a
  statistically same type aggregation and diffusion mechanism.} 
\label{fig:no3}
\end{figure}

It is obvious that the aggregation and diffusion mechanism of the $1
\times 1$ particles follow the RDSR mechanism \cite{f}. The $2 \times
1$ particles also follows the RDSR mechanism except the height
increment of the column at which it finally sticks is of two units
rather one as that of $1 \times 1$ particles. Similar rules are also
valid for the relaxation mechanism of the particles of sizes $3 \times
1$, $4 \times 1$, $....$. But the case is quite different for the $1
\times 2$, $2 \times 2$, $1 \times 3$, $.....$ particles. The
mechanism of the diffusion of these particles is not that of RDSR
type. The growth rules are straightforward generalization of the RDSR
model for extended particles. It seems that the diffusion of each of
such type of particles will cause different type of surface
morphology. For clarification, we show in  Fig.\ref{fig:no2} and
Fig.\ref{fig:no3} schematically the aggregation and diffusion
mechanism for the $1 \times 2$ and $1 \times 3$ particles
respectively. In both of these two figures, the one of the extreme
cells (I and II) corresponding to the extended particles is chosen by
a site (A) of the substrate lattice randomly.

\begin{figure}[ht]
\begin{center}
\rotatebox{-90}
{\epsfxsize=7 cm \epsfysize= 9.5cm \epsfbox{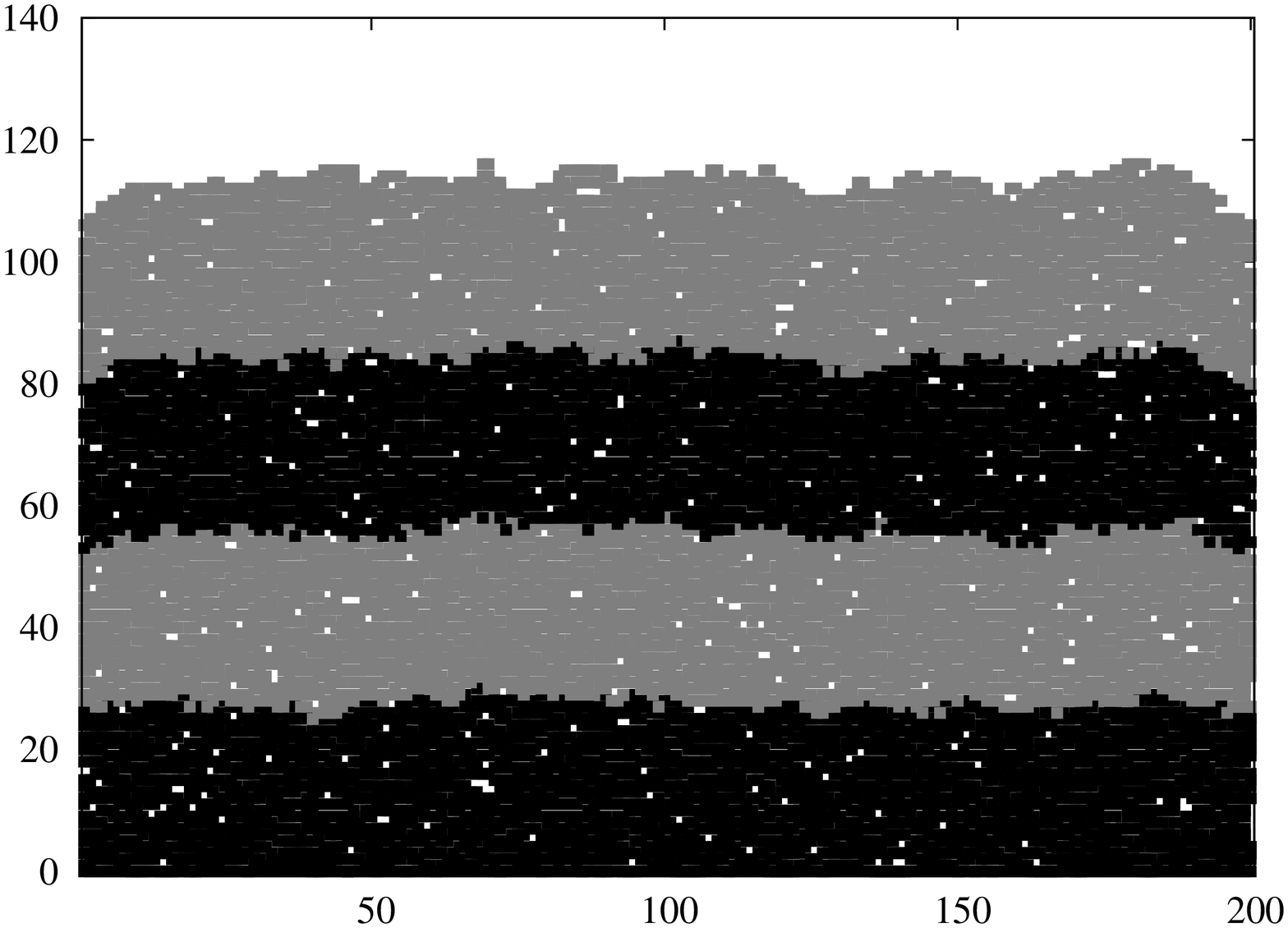}}
\end{center}
\caption{ The rough surface formed due to the deposition of $1 \times 2$
  particles. The total number of particles involved here, have been isolated
  into four set with different color shading showing the change in roughness.}
\label{fig:no3a}
\end{figure}

With these rules, we have developed a  model in $(1 + 1)$ dimension with the
participating particles having different sizes. The morphology of the surface
should be determined by the particles dominating the growth process. During
the growth process bulk defects are allowed to form. But the growth rules have
been set in such a way that the voids formed in the system are closed, so that
possible overhangs should be avoided. The lateral growth property breaks the
up-down symmetry, resulting the presence of non-linearity due to the local
slope ($\nabla h (\vec {\bf r},t)$) fluctuation in the continuum description
of the model. It was previously reported \cite{mj} that the participation of
the $2 \times 1$ particles in the growth process does not significantly change
the morphology of the surface generated from RDSR
mechanism. Moreover, the $1 \times 2$ particles
  create voids in the system during surface growth, so KPZ type of
  growth characteristics may occur. Also the up-down symmetry is
  broken due to the involvement of the $1 \times 2$ particles in the
  growth process. So, the scaling properties of the kinetic roughening of the surface are expected to be modified by the non-linearity introduced by $1 \times 2$ particles.
Due to the above mentioned reasons
 we have simulated the surface and analyzed its kinetic roughening for
the surface formed due to the $1 \times 2$ particles only. The morphology of
the surface formed due to the deposition of $1 \times 2$ particles for the
system size $L = 200$ at time $2 \times 10^2$ is shown in Fig.\ref{fig:no3a}
with four equal time zones. The increment of roughness of the surface
with time is evident from the figure.

The model has been simulated in the lattices of lengths $L=50 \times
2^n$ ($n=0$ to $n=6$). Deposition time is taken as $10^4 \le t \le
10^7$ depending upon the lattice size. The internal structure of the
surface is characterized by `intrinsic width'\cite{wk}. The probable
origin of intrinsic width is the voids, overhangs and large local
slopes. To minimize the effects from intrinsic width, we have
incorporated the noise reduction technique \cite{wk}. The noise
reduction parameter is the number of attempts per site for the actual
aggregation process. In this model, the noise reduction parameter
($m$) was set fixed as $m=10$. With this value of m the surface
morphology shows a stable scaling behavior with repeated independent
simulations. Uniformly distributed uncorrelated noise has been
taken. Depending upon the system size, the results were averaged over
$100$ to $10$ independent runs. Simulations were done on an IBM Server
PC with two 64-bit quad-core POWER5+ processors.

\section{Non-universal scaling of kinetic roughening of the surfaces }

\begin{figure}[ht]
\begin{center}
\rotatebox{-90}
{\epsfxsize=9 cm \epsfysize=9.5 cm \epsfbox{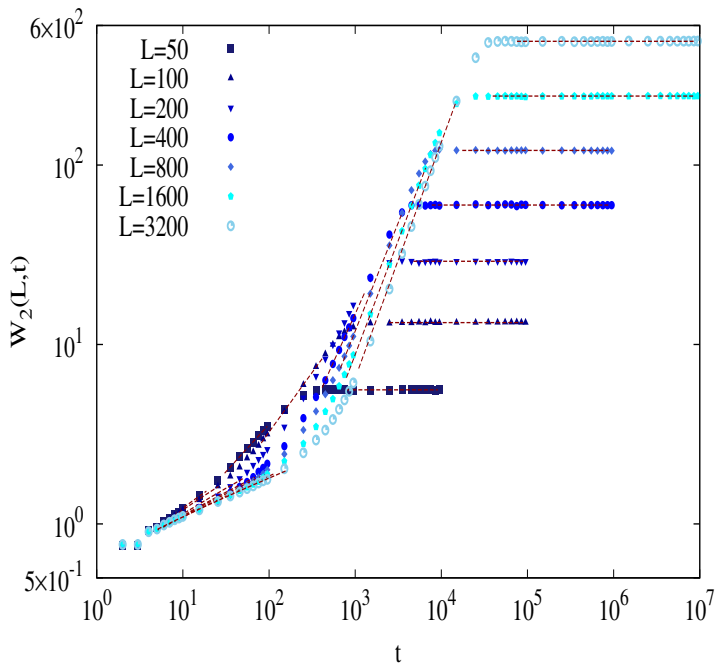}}
\end{center}
\caption{ Log-log plot of $W_2(L,t)$, the 2nd order moment of the
  height fluctuation as defined from equation (\ref {eq:no1}), versus
  $ t$. The dotted lines are the power law fitted according to the
  equation (\ref{eq:no3a}) in three distinct regimes.}
\label{fig:no4}
\end{figure}

The kinetic roughening is characterized by the scaling exponents
corresponding to the moments of height fluctuation defined in equation
(\ref{eq:no1}). The log-log plot of the 2nd order moment of the height
fluctuation for the values of lattice size $50 \le L \le 3200$ is
shown in Fig.\ref{fig:no4}. It is clear from the figure that three
distinct scaling regimes exist, which are separated by two time scales
namely $t_{\times1}$ and $t_{\times2}$. The situation occurred around
the crossover time scale $t_{\times1}$ should not be confused with the
situation occurred when the finite size scaling is affected by the
intrinsic width of the system. Because intrinsic width is a system
size independent effect \cite{wk1}. The crossover between the
intrinsic width affected regime with the unaffected regime is
independent of the system size. But from Fig.\ref{fig:no4} it is clear
that the crossover $t_{\times1}$ is system size dependent. The scaling
form of the time scale $t_{\times 1}$ will be shown latter.

\begin{figure}[ht]
\begin{center}
\rotatebox{-90}
{\epsfxsize=9cm \epsfysize=9.5cm \epsfbox{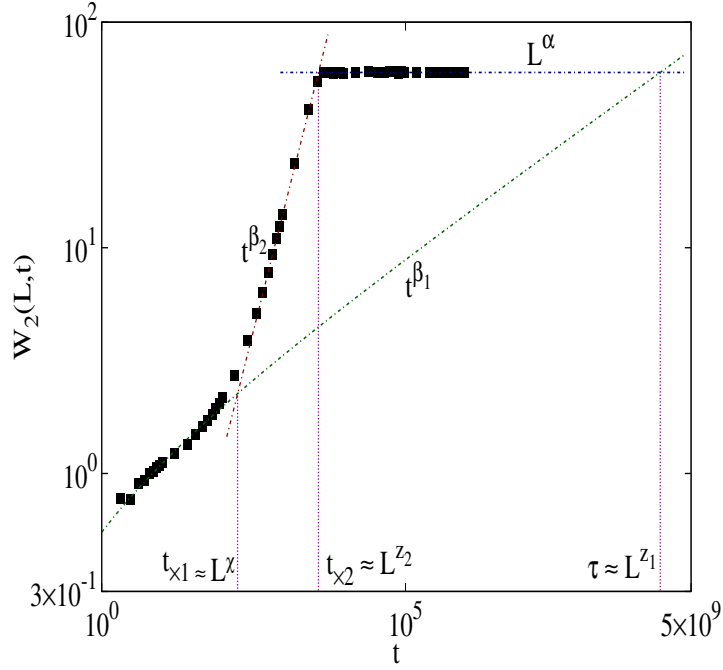}}
\end{center}
\caption{ Representation of different scaling regimes of the 2nd order
  height fluctuation for a particular system size L=400 with the
  respective power law fitting. This fit shows $\alpha = 0.698 \pm
  0.002$, $\beta_1=0.22 \pm 0.02$ and $\beta_2=1.21 \pm 0.04$.}
\label{fig:no5}
\end{figure}

To demonstrate the scaling behavior of the rough surface evolved from the
present discrete model, the time evolution of the height fluctuation
corresponding to a particular system size ($L=400$) has been shown in
Fig.\ref{fig:no5}. The nature of the plot in Fig.\ref{fig:no5} is quite
similar to that of the plot for the competitive growth model consisting of
RDSR and BD \cite{cr}. But the situation is different here. Only one kind of
particles are taking part in the growth process. So, no other time scale can
appear depending upon the abundances of the different kinds of particles
participating in the growth process as in the discrete model
  described in ref \cite{cr}. Thus, the scaling behavior for the present
model can not be defined by the equation (\ref{eq:no3}). Guided by the nature
of the scaling of a growth process \cite{ba}, three power laws with different
exponents have been fitted in three different scaling regimes as follows

\bea
\label{eq:no3a}
W(L,t) \sim t^{\beta_1} \;\;\;\;\;\;\;\;\;\;\;\;\;\;\;\;\; t \ll t_{\times1} \n \\
W(L,t) \sim t^{\beta_2} \;\;\;\;\;\;\; t_{\times 1} \ll t \ll t_{\times 2} \n \\
W(L,t) \sim L^\alpha  \;\;\;\;\;\;\;\;\;\;\;\;\;\;\;\;\; t \gg t_{\times2}
\eea

Also, in Fig.\ref{fig:no4} the power law fitting for different system
sizes has been shown by dotted lines. The scaling behavior shown in
equation (\ref{eq:no3a}) indicates that, though the surface evolved
with a definite mechanism, two different growth processes belonging to
different universality classes are governing the overall kinetic
roughening process. Such a scaling behavior of a growth model has not
been noticed previously in literature. The system contains three
characteristic time scales $t_{\times1}$, $\tau$ and $t_{\times2}$
(see Fig.\ref{fig:no5}). The time scale $\tau$ denotes the saturation
time for a growth process belonging to a certain universality class
with the growth exponents $\alpha$ and $\beta_1$. Another growth
process belonging to different universality class seems to occur
beyond $t \sim t_{\times1}$ with the growth exponents $\alpha$ and
$\beta_2$. From Fig.\ref{fig:no5}, it is seen that the two independent
growth processes having saturation time scales $\tau$ and $t_{\times
  2}$ actually determine the crossover time scale $t_{\times 1}$. So,
the crossover time scale $t_{\times1}$ should depend on the time
scales $\tau$ and $t_{\times2}$. The crucial time scale $t_{\times1}$
is the crossover between these two growth processes. The saturated
rough surface is characterized by the latter growth process dominating
beyond $t_{\times1}$. The time scales corresponding to the two
distinguished growth processes should behave as
\bea
\label{eq:no3b}
\tau \sim L^{z_1} \n \\
t_{\times2} \sim L^{z_2}
\eea
Also from the nature of the height fluctuation (see Fig.\ref{fig:no4})
it can be argued that the crossover time scale $t_{\times1}$ can be
scaled with the system size $L$ as
\be
\label{eq:no3c}
t_{\times1} \sim L^\chi
\ee
Where $z_1=\alpha/\beta_1$ and $z_2=\alpha/\beta_2$ are the two
dynamic exponents corresponding to the two growth processes occurring
in two different time regimes, $\chi$ is a different scaling
exponent. Since the crossover time scale $t_{\times 1}$ is dependent
on the other two independent time scales $\tau$ and $t_{\times1}$ so
the scaling exponents  $\chi$ depends naturally on the two independent
exponents $z_1$ and $z_2$.

From the scaling behavior, we argue the simplest possibility that the
kinetic roughening of the surface is occurring by the two growth
processes in two different time regimes (viz $t \ll t_{\times1}$ and
$t \gg t_{\times1}$) independently. With this consideration, we
propose that the overall morphology of the surface is governed by the
scaling relation which can be represented as the linear sum of two
scaling functions corresponding to each of the growth process
dominating in different regimes, leading to the saturated rough
surface having unique roughness exponent. So, mathematically the
proposed scaling relation can be represented as

\be
\label{eq:no4}
W_2(L,t) \sim L^\alpha \left [f_1\left (\frac{t}{L^{z_1}} \right )+f_2
  \left (\frac{t}{L^{z_2}} \right )\right ]
\ee
where the various scaling functions are defined by\\

\bea
\label{eq:no4a}
f_1(u_1) \sim u_1^{\beta_1} \;\;\;\; \;\;\;\;\;\;\;\; \;\;\;\;\;\;\;\; \;\;
u_1 \ll \frac{1}{L^{z_1 - \chi}} \n \\
f_1(u_1) \sim constant \;\;\;\;\;\;\;\;\; \;\;\;\;\; u_1 \gg
\frac{1}{L^{z_1-\chi}} \n \\
f_2(u_2) \sim constant \;\;\;\; \;\;\;\;\;\;\;\;\;\;  u_2 \ll
\frac{1}{L^{z_2-\chi}} \n \\
f _2(u_2) \sim u_2^{\beta_2} \;\;\;\;\;\;\;\;\;\;\;\;\;\;\;\;
\frac{1}{L^{z_2-\chi}} \ll u_2 \ll 1 \n \\
f_2(u_2) \sim constant \;\;\; \;\;\;\;\;\;\;\;\;\;\;\;\;\;\;\;\;\;\;\;\; u_2 \gg 1
\eea

To observe the appropriate scaling and the crossover,
  we proceed as follows. According to the scaling relation (\ref{eq:no4}), in
  the time regime $t \ll t_{\times1}$ i.e, for $t/L^{z_1} \ll
  1/L^{z_1-\chi}$ the scaling functions $f_1$ and $f_2$ will be
\bea
f_1\left( \frac{t}{L^{z_1}} \right) \sim
\left(\frac{t}{L^{z_1}}\right)^{\beta_1}  \n  \\
f_2\left(\frac{t}{L^{z_2}} \right) \sim constant \n
\eea
Thus from equation (\ref{eq:no4}) the scaling relation becomes
\be
\label{eq:no4b}
W_2(L,t)/L^\alpha \sim \left(\frac{t}{L^{z_1}} \right)^{\beta_1} \hspace{1cm} when
\hspace{0.3cm} \frac{t}{L^{z_1}} \ll
\frac{1}{L^{z_1-\chi}}
\ee
\\

\begin{figure}[ht]
\begin{center}
\rotatebox{-90}
{\epsfxsize=9cm \epsfysize=9.5cm \epsfbox{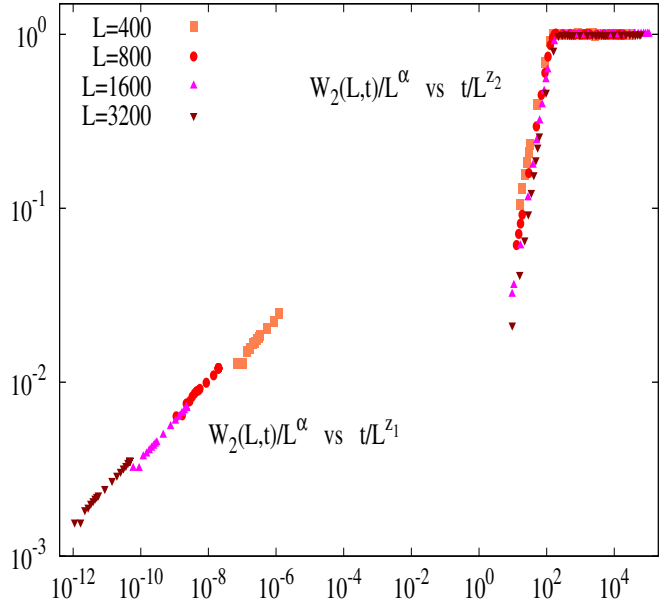}}
\end{center}
\caption{ The scaling plot in log-log scale shows that
  $W_2(L,t)/L^\alpha$ is plotted against $t/L^{z_1}$ for $t \ll
  L^{z_1}$ and $W_2(L,t)/L^\alpha$ is plotted against $t/L^{z_2}$ for
  $t \gg L^{z_1}$. As predicted from the scaling relation given by the
  equation (\ref{eq:no4}), two distinct scaling functions given by
  equation (\ref{eq:no4a}) are seen to exist prominently.}
\label{fig:no6}
\end{figure}

For the time regime $t_{\times1} \ll t \ll
  t_{\times2}$ i.e, for $1/L^{z_2- \chi} \ll t/L^{z_2} \ll 1$ the
  scaling functions $f_1$ and $f_2$ will looks like
\bea
f_1\left(\frac{t}{L^{z_1}}\right) \sim constant \n \\
f_2\left(\frac{t}{L^{z_2}}\right) \sim
\left(\frac{t}{L^{z_2}}\right)^{\beta_2} \n
\eea
So according to the equation (\ref{eq:no4}) the
  scaling relation reduces to \\
\be
\label{eq:no4c}
W_2(L,t)/L^\alpha \sim
\left(\frac{t}{L^{z_2}}\right)^{\beta_2} \hspace{1cm} when \hspace{0.3cm}
\frac{1}{L^{z_2-\chi}} \ll \frac{t}{L^{z_2}} \ll 1
\ee 
\\
The time regime $t \gg t_{\times2}$ i.e, when
  $t/L^{z_2} \gg 1$ the scaling  functions $f_1$ and $f_2$ behave like
\bea
f_1\left(\frac{t}{L^{z_1}}\right) \sim constant \n \\
f_2\left(\frac{t}{L^{z_2}}\right) \sim constant \n
\eea
In this time regime the scaling relation turns out as
\be
\label{eq:no4d}
W_2(L,t)/L^\alpha \sim constant  \hspace{1cm} when \hspace{0.3cm}
\frac{t}{L^{z_2}} \gg 1
\ee
\\
The above scaling relations in equations (\ref{eq:no4b}), (\ref{eq:no4c}) and (\ref{eq:no4d}) satisfy the
observation in equation (\ref{eq:no3a}). The complete data collapse in the log-log plot of $W_2(L,t)/L^\alpha$ in different time regimes as shown the Fig.\ref{fig:no6} also confirm the above scaling relation.

 Thus, for this present discrete model, the kinetic roughening of the
 surface can not be scaled with a unique scaling function having two
 independent scaling exponents. It thus deviates from the universal
 nature of the scaling of the kinetic roughening of the surfaces.

The visualization of the scaling relation defined in equation
(\ref{eq:no4}) with the scaling functions given in equation
(\ref{eq:no4a}), is shown in Fig.\ref{fig:no6} with the data collapse
in two different time regimes $t \ll t_{\times1}$ and $t \gg
t_{\times1}$. The scaling functions have been plotted for different
set of $L$, $\alpha$, $\beta_1$ and $\beta_2$ values, because of the
strong finite size effect on $\alpha$ and weak finite size effect on
$\beta_1$ and $\beta_2$. Though the x-coordinate for each plot are
different (one is $t/L^{z_1}$ and the other  $t/L^{z_2}$), we plot in
the same graph to show the existence of two distinguished scaling
regimes $t \ll t_{\times1}$  and $t \gg t_{\times1}$. The large gap
between the two scaling regimes, $t \ll t_{\times1}$ and $t \gg
t_{\times1}$, is due to the large difference of two dynamic exponents
$z_1$ and $z_2$. Also, long crossover region around $t \sim t_{\times
  1}$  is one of the reason for such a large gap. The data collapse
for scaling functions defined in equation (\ref{eq:no4a}) with the
exponents defined in equation (\ref{eq:no3a}) for the respective time
regimes defined in equations (\ref{eq:no3b}) and (\ref{eq:no3c}),
confirms the scaling relation defined by the equation (\ref{eq:no4}).

The scaling behavior shown in Fig.\ref{fig:no4} can be explained
physically in the following way. In our model, the particles diffuse
along the surface in search of a stable position. During the growth
process, there is a interplay of  diffusion with the finding the
stable position for each particle. Initially, since the total number
of involved particles in the growth process is small, each particle is
allowed to diffuse throughout the surface. So, within this time,
diffusion of particles is dominated over the process of finding the
stable position. After the relaxation, when a $1 \times 2$ particle
sticks to the substrate lattice, it implies that the constituent two
$1 \times 1$ particles stick with the same height. Thus, from the $1
\times 1$ particle point of view the height-height fluctuation also
decreases. Due to the above two reasons the overall height-height
fluctuation of the surface during this time period will be very small,
resulting a small value of the dynamic exponent ($\beta$). However
with the increase of time, more and more particles do take part in the
growth process. Due to the relaxation rules there is a fair
probability of getting a stable position for each particle within the
short-range sites. Effectively, the particles are restricted to the
connected sites of the lattice site on which they were deposited. In
other words, the particles are now localized almost stopping the
long-range relaxation. Thus, an instability occurs due to the piling
of particles on the upper terraces and restriction of the relaxation
on the lower terraces. This increases the height-height fluctuation,
resulting a large value of the dynamic exponent ($\beta$). This effect
is quite similar to the Ehrlich-Schwoebel (ES) effect
\cite{es1,es2,s,kr,p} which arises due the presence of a ES
barrier. This effect induces an instability by hindering step-edge
atoms on upper terraces from going down to lower terraces in the MBE
type of growth processes. A discrete solid-on-solid model \cite{bk} of
epitaxial growth without bulk defect was proposed, which takes into
account the ES effect with the introduction of a parameter ($R_{inc}$)
called the incorporation radius as a ES barrier. Another reasonably
realistic bulk defect induced discrete model \cite{sk} of epitaxial
growth in ($1+1$) dimension was presented in such a way so that the
kinetic roughening is controlled by the interplay of the mound
instability with the KPZ roughening. In this model, the diffusion of
the particles were partially controlled by the parameter $E$, which
actually selects the direction of diffusion with a probability $\exp
(-E)$. In both of the above models the instability occurs due to the
presence of a step edge barrier, in former case it is infinite while
in the latter case it is finite. However, in our model, we do not put
any step edge barrier explicitly, the instability occurred here is
completely self organized.

The complicated scaling behavior of the rough surface evolved from the
present discrete model is represented with a linear sum of two
independent scaling functions corresponding to different growth
processes. Now we point out the unusual behavior of the scaling
function around $t \sim t_{\times1}$. The crossover around
$t_{\times1}$ leads to a morphological `phase transition' from one
universality class to another universality class. Such a morphological
phase transition with a unique growth mechanism has not been observed
previously for any discrete growth model. In this context, we may also
mention that another morphological linear-nonlinear `phase transition'
is seen to occur around a critical probability of deposition of the $1
\times 2$ particles, beyond a characteristic length scale, for a
competitive growth model involving particles of sizes $1 \times 1$, $2
\times1$ and $1 \times 2$ \cite{mj}. Below we would like to study the
variation of the roughness exponents with the system size L.

\section{Finite size effect on roughness exponent}

The earlier work by Krug and Meakin \cite{km} had shown that for a
nonlinear KPZ growth model the roughness exponents were affected by
the finite size dependence of the steady state growth rate of the
system. The demonstration of the finite size effect on the growth rate
of the evolution of the surface is shown first. The growth rate of the
surface is defined as $V(L,t)=\frac{d{\langle h(t)
    \rangle}}{dt}$. According to the suggestion by Krug and Meakin
\cite{km}, the steady state growth rate is scaled with the system size
$L$ as

\be
\label{eq:no5}
V(L,t \rightarrow \infty)=V(L \rightarrow \infty) - \Lambda L^{-\nu}
\ee

\begin{figure}[ht]
\begin{center}
\rotatebox{-90}
{\epsfxsize=7cm \epsfysize=7cm \epsfbox{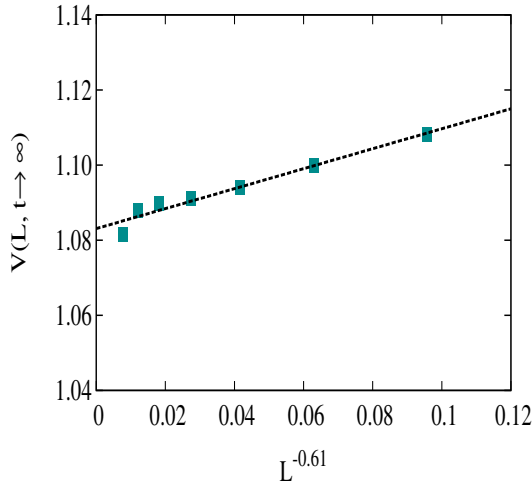}}
\end{center}
\caption{ Fitting of steady state velocity in the asymptotic limit
  versus $L^{-\nu}$ with $\nu=0.61 \pm 0.05$.}
\label{fig:no7}
\end{figure}

The scaling behavior of the velocity $V$ in equation (\ref{eq:no5}) is
shown in Fig.\ref{fig:no7}. The fitting shows $\nu = 0.61 \pm 0.05$,
with asymptotic limit of the steady state velocity is $V(L \rightarrow
\infty)=1.083 \pm 0.02$.  It was predicted \cite{km} that the
roughness exponent for large enough system for growth models belongs
to the KPZ universality class, would be related with $\nu$ as
\be
\label{eq:no6}
\alpha = 1-\nu/2
\ee

Strong finite size effect on $\alpha$ is also observed in the present
model. The following scaling relation for the finite size dependence
on $\alpha$ has been suggested \cite{r} earlier.

\be
\label{eq:no7}
\alpha(L)=\alpha(L \rightarrow \infty)+\Upsilon L^{-\delta}
\ee

\begin{figure}[ht]
\begin{center}
\rotatebox{-90}
{\epsfxsize=7cm \epsfysize=7cm \epsfbox{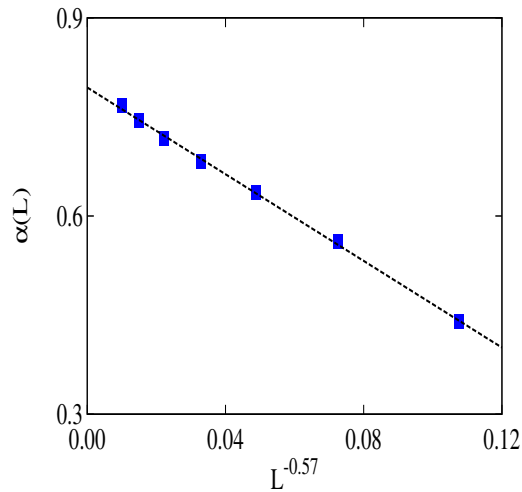}}
\end{center}
\caption{ Finite size dependence of the roughness exponent
  $\alpha$. The fitting of $\alpha$ with $L^{-\delta}$ shows that
  $\delta=0.57 \pm 0.03$.}
\label{fig:no8}
\end{figure}

In Fig.\ref{fig:no8}, the $\alpha$ values are fitted according to the
above scaling relation given by equation (\ref{eq:no7}) with
$\delta=0.57 \pm 0.03$. For asymptotically infinite system the
roughness exponent $\alpha(L \rightarrow \infty) = 0.794 \pm 0.005$
which is comparable with the prediction in equation
(\ref{eq:no6}). The relation between the exponents shown in equation
(\ref{eq:no6}) is based on the realization that the Family-Vicsek
scaling relation (equation (\ref{eq:no2})) is satisfied. In the
hydrodynamic limit, the present model satisfies the scaling relation
given in equation (\ref{eq:no4}) rather than in equation
(\ref{eq:no2}). So, it can be argued that the correction due to the
finite size dependence of the roughness exponent will be not like that
of the model which follows KPZ type of growth. We have compared these
two models because, both of these models have inclination dependent
current due to the lateral growth property, which is the source of KPZ
like behavior in the growth process. Growth models with power law
distributed noise events $P(\eta(\vec{\bf r},t)) \sim \eta^{-(1+\mu)}$
show such values of $\alpha$ \cite{bhk,ls} with different values of
$\mu$. However, in our present model, the noise distribution is
$\delta$-correlated with uniform amplitude. Rare events are not
occurring here. With the time evolution of the system, a multifractal
behavior is also seen to occur in the present model \cite{mj}, similar
in nature as that of the rare event dominated growth model
\cite{bbj}.

The values of the two dynamic exponents $\beta_1$ and $\beta_2$ remains steady
for large system size. The values of this two dynamic exponents are found to
be $\beta_1=0.22 \pm 0.03$ and $\beta_2=1.22 \pm 0.01$ for large enough
systems. Such a large value of the dynamic exponent
  $\beta_2$ still has not been observed in literature. Due to strong
finite size dependence of $\alpha$ and weak finite size
effect on the values of $\beta_1$ and $\beta_2$, the crossover time scale
$t_{\times1}$ dependent on system size $L$ with a exponent $\chi$ as given in
equation (\ref{eq:no3c}). Moreover, $\alpha + z_1 \ne 2$ and $\alpha+z_2 \ne
2$; this immediately points out the breakdown of the Galilean invariance of
the system. It is to be noted that the value of dynamic exponent $\beta_1$ is
very close to that of the EW model with $\beta_1=\frac{1}{4}$. Above the
system dependent time scale $t \sim t_{\times 1}$ the dynamic exponent
$\beta_2$ has a very high value. The transition of the dynamic
exponent from low value to a higher value can be visualized by looking
critically to the individual configuration of the surface and the
bulk. As per the aggregation and diffusion rules, it appears that
initially the surface moves compactly without having voids and that is
why the dynamic exponent ($\beta_1$) in that region is close to that
of the RDSR model. In latter time, voids are  incorporated into the
system and an ES like instability is found to occur. This self
organized ES effect triggers the rapid roughening of the surface. The
dynamic exponent ($\beta_2$) becomes high in that region. The
experimental observation of rapid roughening involving extended
particles in the growth of organic thin film (Diindenoperylene) was
reported previously \cite{dsr}.

\section{ Bulk properties and its scaling}

To have a deep insight into the internal structure of the interface,
the bulk properties of the system are of great interest. The diffusion
mechanism for the present growth model is unique by its
definition. Closed voids created due to such diffusion mechanism made
the system porous in its own way. So, the bulk property will be
different from the other porous systems, created from different kind
of aggregation and diffusion mechanism. The bulk properties of a
system can be quantified with the definition of porosity. Since the
number of closed voids can be determined accurately, we define,
porosity P for this particular system in a similar way as defined in
Ref \cite{sr}.

\be
\label{eq:no8}
P= N_v/N_t
\ee

Where $N_v=$ Number of voids and $N_t=$ Number of voids + Number of
particles deposited. A deposition process where the number of
particles is conserved, is characterized by the particle flux $J$. As
shown by Krug \cite{krug}, the deposit density $\rho$ can be related
with the growth velocity $v$ as

\be
\label{eq:no8a}
\rho=J/v
\ee

 According to the equation (\ref{eq:no8a}) the deposit density, which
 actually characterized by the quantity porosity, will be scaled with
 the system size as that of the velocity with the modification due to
 the particle flux dependence on the system size in the asymptotic
 limit. To see the asymptotic limit of the porous structure of the
 system at saturation with system size, we propose a scaling relation
 of the porosity as

\be
\label{eq:no9}
P(L,t \rightarrow \infty)=P(L \rightarrow \infty) + \Gamma L^{-\eta}
\ee

The above proposed scaling relation is new in the present literature.

\begin{figure}[ht]
\begin{center}
\rotatebox{-90}
{\epsfxsize=7cm \epsfysize=7cm \epsfbox{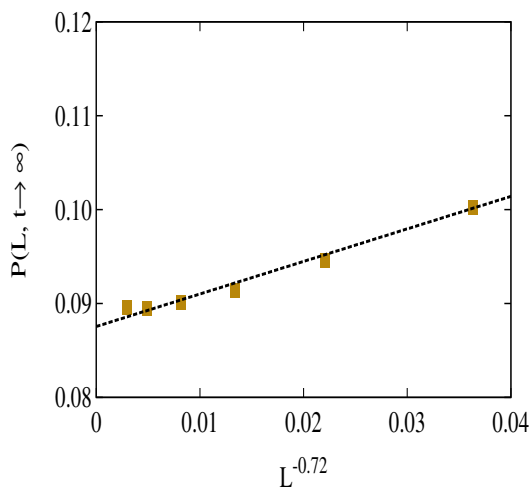}}
\end{center}
\caption{ Plot of porosity in the asymptotic limit with $L^{-\eta}$
  with the exponent $\eta=0.72 \pm 0.02$. }
\label{fig:no9}
\end{figure}

From physical point view, porosity can be defined as the reverse of
the deposit density. That is 
\be
\label{eq:no10}
P=1-\rho
\ee
From equation (\ref{eq:no8a}) for uniform flux $J$
\bea
\label{eq:no11}
1-P=J/v \n \\
P \sim v-1 \\ \n
\eea
Fig.\ref{fig:no9} shows the behavior of the porosity at the asymptotic
limit with different system sizes. It shows that asymptotic value of
the porosity $P(L \rightarrow \infty)$ as $0.087 \pm 0.005$ and the
coefficient $\Gamma=0.34 \pm 0.013$. The prediction from equation
(\ref {eq:no11}) is well in agreement with the results shown in
Fig.\ref{fig:no7} and Fig.\ref{fig:no9}.

\section{conclusion}

The kinetic roughening of the surface created due to a nonlinear
discrete growth model, is studied here. Several features, not
previously observed corresponding to kinetic roughening, are observed
in the present model in (1+1) dimension. To summarize, we mention the
following points systematically. The finite size scaling of the rough
surface shows a different type scaling nature. Two distinguished time
scales, corresponding to the height fluctuation, emerge in the system
in (1+1) dimension. They separate three scaling regimes with different
scaling exponents as well as scaling functions. To characterize this
kinetic roughening, a new scaling relation is proposed, which is
represented as the linear sum of two scaling functions valid for two
distinguished scaling regimes. The existence of two scaling regimes
with small and large values of dynamic exponent $\beta$ is well
explained with the occurrence of a self organized
Ehrlich-Schwoebel (ES) like
instability (caused due to the localization of the extended
particles) which triggers the rapid roughening of
  the surface. Due to the finite size effect on the growth rate, the
scaling exponents are also affected by the finite size of the
system. The finite size effect on the roughness exponent is scaled
with a scaling relation. The scaling exponent for this scaling
relation is well compared with the prediction made by Krug and
Meakin. The bulk nature of the system for different sizes is shown
through a new scaling relation.

\section{acknowledgment}

One of the authors (P.K.M.) would like to thank the University Grant
Commission (UGC), New Delhi, India for financial support to carry out this
work. We are grateful to DST-FIST, New Delhi, India for providing us the
necessary help. We are indebted to anonymous referees for critical suggestions
and comments.

\end{document}